\documentclass[reprint, prb, amsmath, amssymb, ]{revtex4-2} 			 % 2-column prb

\usepackage[utf8]{inputenc}
\usepackage[english]{babel}

\usepackage{color}
\usepackage{graphicx}
\usepackage{bm} %bold math
\usepackage{dcolumn} % Align table columns on decimal point
\usepackage[colorlinks=true, citecolor=blue, linkcolor=blue]{hyperref}%

\allowdisplaybreaks

\begin{document}

\title{Exciton dynamics in CdTe/CdZnTe quantum well} %Title of paper

\author{A. V. Mikhailov}
\email{mikhailovav@yandex.ru}
\author{A. S. Kurdyubov}
\author{E. S. Khramtsov}
\author{I. V. Ignatiev}
\affiliation{Spin Optics Laboratory, St. Petersburg State University, Ulyanovskaya 1, Peterhof, St. Petersburg, 198504, Russia}
\author{B. F. Gribakin}
\author{S. Cronenberger}
\author{D. Scalbert}
\author{M. R. Vladimirova}
\affiliation{Laboratoire Charles Coulomb, UMR 5221 CNRS, Univ. Montpellier, Place Eug{\'e}ne Bataillon, F-34095, Montpellier, France}
\author{R. André}
\affiliation{Univ. Grenoble Alpes, CNRS, Institut N{\'{e}}el, 38000 Grenoble, France}

\date{\today}

\begin{abstract}
Exciton energy structure and population dynamics in a wide CdTe/CdZnTe quantum well are studied by spectrally-resolved pump-probe spectroscopy. %
 Multiple excitonic resonances in reflectance spectra are observed and  identified by solving numerically three-dimensional Schr{\"{o}}dinger equation.
The pump-probe reflectivity signal is shown to be dominated by the photoinduced  nonradiative broadening of the excitonic resonances, while pump-induced exciton energy shift and reduction of the oscillator strength appear to be negligible. 
This broadening is induced by  the reservoir of dark excitons with large in-plane wave vector, which  are coupled to the the bright excitons states.
The dynamics of the pump-induced nonradiative broadening observed experimentally is characterised by three components:  signal build up on the scale of tens of picoseconds  (i) and bi-exponential  decay on the scale of one nanosecond (ii) and ten nanosecons (iii).
Possible mechanisms of the reservoir population and depletion responsible for this behaviour are discussed. 
\end{abstract}
\maketitle %\maketitle must follow title, authors, abstract and \pacs

\section*{Introduction}

Excitons in quantum wells (QWs) demonstrate remarkable optical properties. In high-quality structures, they give rise to intense photoluminescence (PL) lines under nonresonant optical excitation, as well as strong features in  reflection spectra. These effects are associated with a violation of the wave vector selection rules for optical transitions in the presence of quantum well/barrier layer interfaces. However, for excitons propagating in the QW plane, the in-plane wave vector is a good quantum number. Correspondingly, when exciton wave vector ($K_x$) exceeds the wave vector of light ($K_c$) in the QW layer  excitons can not emit or absorb photons. These excitons are usually referred to as nonradiative, or dark excitons. 

In extensively studied high-quality GaAs-based QWs, the life time of dark excitons  has been shown to reach tens of nanoseconds~\cite{Trifonov-PRB2015, Kurdyubov-PRB2021, Kurdyubov-PRB2023}. This is much longer than the radiative lifetime of bright excitons with $K_x<K_c$, which is typically $1000$ times shorter \cite{Sermage-LJP1993,Andreani1991}. Correspondingly,  dark excitons accumulate in a so-called nonradiative reservoir, up to densities orders of magnitude exceeding the density of bright excitons. This reservoir strongly affects optical properties as well as the dynamics of bright excitons.  

In this work, we  study energy spectrum and dynamics of excitons in a wide ($L=47$~nm) CdTe/Cd$_{1-x}$Zn$_{x}$Te QW with small content of Zn in the barrier layers, $x = 0.05$. This is a type-I structure for heavy-hole excitons (an electron and a  heavy hole are located in the same layer) and type-II for light hole ones (an electron is located in the QW and a light hole in the barrier layers)~\cite{Mariette-PRB1988, Merle-JCG1990, Tuffigo-PRB1991, Peyla-PRB1992, Abdi-Ben-JAlloyC}. Therefore the dynamics of heavy-hole excitons in this QW may differ from that in GaAs QWs.

Exciton dynamics in CdTe/Cd$_{1-x}$Zn$_{x}$Te QW is addressed by time-resolved photoinduced reflection spectroscopy. It appears that  the main effect induced by resonant optical pumping is the nonradiative broadening of exciton resonances, while pump-induced exciton energy shift and reduction of the oscillator strength are  negligibly small. We  interpret this broadening as being due to the interaction of photo-created excitons with other quasi-particles in the system such as free carriers, other excitons, trions, phonons. 

Scattering of the photo-created excitons on each other can only contribute at very short time ($< 10$~ps) due to rapid exciton recombination.  These also include pure dephasing (optical decoherence) on the ~10 ps scale usually addressed in four wave mixing experiments~\cite{Honold-PRB1989, Mayer-PRB1995, Portella-Oberli-PRB2002}. Here we are interested in the remaining mechanism, the exciton scattering involving  the reservoir of dark excitons that take place on the scale of $1-10$~ns. The latter process also involves acoustic phonons. This is a kind of inelastic scattering, its efficiency depends on the temperature (phonon density) and the dark reservoir density. 

\section{Optical characterization}
\label{experimental}
The structure under study was grown by the molecular beam epitaxy (MBE) on a Cd$_{0.96}$Zn$_{0.04}$Te substrate with a crystallographic orientation (001). It contains a 47 nm-wide CdTe QW layer sandwiched between Cd$_{0.95}$Zn$_{0.05}$Te barrier layers. The thickness of the bottom barrier layer, $L_b = 1064$~nm, the top one, $L_t = 102$~nm. The thickness of all the layers are precisely controlled (with accuracy 1\%) during the growth procedure. The gradient of the thicknesses is also very small ($<1$~\%). The thickness of the top barrier layer was chosen to get the constructive interference of the light waves reflected from the sample surface and the QW layer. Due to this constructive interference, the lowest exciton resonance in the reflectance spectrum is observed as a peak-like feature. Because of large lattice mismatch between the CdTe and ZnTe crystals (6\%~\cite{Tatarenko-APL1990}), the strain-induced splitting of the heavy-hole and light hole states occurs in the structure. As a result, the heavy-hole exciton states are split off from  those of the light-hole excitons and can be studied separately.

To get the first insight into the heavy-hole exciton energy spectrum of the structure under study, two standard experimental techniques has been used, namely, PL and reflectance. 
The PL was excited by a continuous-wave (CW) Ti:sapphire laser tuned to the barrier layers absorption band ($E_{\text{exc}} = 1. 653$~eV) and detected by a spectrometer with the 1800 gr./mm grating equipped by a nitrogen-cooled CCD array. 

The reflectance spectra were measured at nearly normal incidence (deviation angle $ < 10^{\circ}$) using  $\approx 80$~fs-long pulses delivered by Ti:sapphire laser at $80$~MHz repetition rate. These spectrally broad pulses allowed us to cover the whole spectral range of interest. The reflected laser beam was detected by the same spectrometer. The normalization of the detected spectrum on the laser beam spectral profile was used to obtain the reflectance spectrum of the sample.  The laser spot diameter on the sample surface is set to $d=100$~$\mu$m.

\begin{figure}
   \includegraphics[width=1\columnwidth]{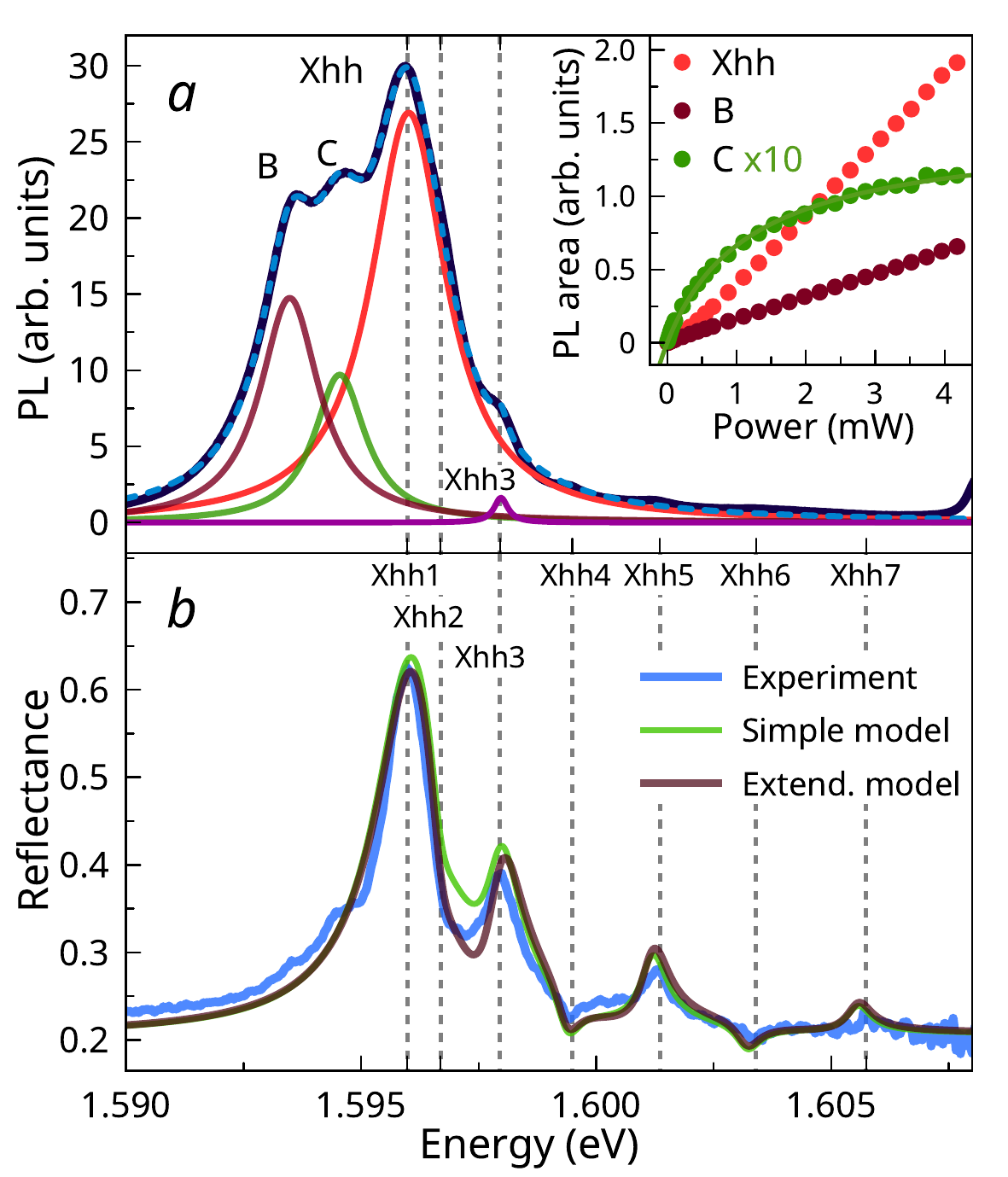}
   \caption{(a) PL spectrum measured at $E_{\text{exc}}=1.653$~eV and $P_{\text{exc}}=0.6$~mW. Laser spot diameter, $d = 0.1$~mm. Dashed curve is a fit to the data with the sum of four Lorentzians (solid lines). Inset shows the power dependence of integral areas of spectral the components labeled as Xhh, B, and C (points). Solid curves are linear approximations for the Xhh and B components and the approximation by function~(\ref{IC}) with $P_0 = 1.2$~mW for the C component. (b) Reflectance spectrum (noisy curve) and its modeling in the framework of the simple and extended models (see the text for details). The calculated spectra are shifted by $2.15$~meV and scaled using the factor 0.93. Vertical dashed lines marked X$_{hh1}$, X$_{hh2}$ etc, indicate the energies of the quantum-confined heavy-hole exciton states. Sample temperature $T = 4$~K for both experiments.}
   \label{Fig1}
\end{figure}

Figure~\ref{Fig1} shows the PL and reflectance spectra of the sample. 
Four Lorenzian components can be identified in the PL spectrum.  The component Xhh can be attributed to PL of the lowest quantum-confined state of the heavy-hole exciton in the CdTe QW. Its integral intensity increases almost linearly  with the excitation power [see the inset in Fig.~\ref{Fig1}(a)], the energy slightly decreases (by about 0.05~meV for $P_{\text{exc}}=5$~mW, not shown in the figure), and the half-width at half-maximum (HWHM) increases from 0.7~meV at small power to 1.2~meV at $P_{\text{exc}}=5$~mW (not shown in the figure). 
Above Xhh line, another emission peak can be distinguished. From comparison with the reflectance spectrum and the modelling (see below) we identify it as Xhh3 quantum-confined exciton state.

Below Xhh line, two emission lines, marked as B and C, are observed. Similar components in the PL spectra of CdTe QWs were observed in Refs.~\cite{Merle-JCG1990, Tuffigo-PRB1991}. The energy splitting between the lines B and Xhh  is of $\approx 2.5$~meV,  close to the binding energy of negative trions in CdTe QWs~\cite{Trion-PRL1993, Ciulin-PRB2000, Portella-Oberli-PRB2002}. The energy and HWHM of the line B increase with increasing power in the same way as for the Xhh. These observations suggest that line B stems from the emission of the heavy-hole trion. 

The energy separation of the C component from the Xhh line is of about 1.5~meV. Its intensity is considerably smaller and grows sublinearly with the excitation power. 
We assume that this component is due to presence of defect centers localizing excitons. Indeed, let the areal density of these centers be $N_c$ and the corresponding bound exciton lifetime be $\tau_c$. The balance equation for the density $n$ of the bound excitons is:
\begin{equation}
\frac{dn}{dt} = P\left(1-\frac{n}{N_c}\right) - \frac{n}{\tau_c}.
\label{EqC}
\end{equation}
Here $P$ is the rate of excitation of the centers and factor $(1-n/N_c)$ describes the probability that the center is not occupied by an exciton. Under  CW excitation, the derivative in Eq.~(\ref{EqC}) is zero and the population $n$ is easily determined. The PL intensity is proportional to $n$:
\begin{equation}
I_c(P) = A n = A \frac{P}{P+P_0},
\label{IC}
\end{equation}
where $A$ is the scaling factor and $P_0 = N_c/\tau_c$. The function describes quite well the power dependence of the component C, see inset in Fig.~\ref{Fig1}. 

Similar behavior of the components is observed under quasiresonant excitation ($E_{\text{exc}}=1.606$~eV). The total PL intensity of all three components depends linearly on the excitation power that indicates the absence of any noticeable concentration of quenching centers.

Reflectance spectrum shown in Fig.~\ref{Fig1}(b) demonstrates a number of optical resonances in the form of peaks and dips. 
There is no noticeable Stokes shift between Xhh1 energy in PL and reflectance spectra. This  indicates high quality of the structure and negligible exciton localization. Multiple resonances in the reflectance spectrum are due to optical transitions associated to different quantum-confined exciton states. Because the width of the QW under study is large compared to the bulk exciton Bohr radius $a_B\simeq5$~nm, several closely located quantum-confined excitons states can be observed above the lowest exciton state. 

Similar quantum-confined states in wide CdTe/CdZnTe QWs were observed in the PL spectra in Ref.~\cite{Tomassini-PRB1995}. The authors analyzed these states in the framework of a center-of-mass quantization model considering the exciton as a single particle, which motion across the QW layer is quantized. Although this model is applicable for wide QWs it contains approximations, which reduce the accuracy of obtained results, in particular, for the exciton-photon coupling~\cite{Khramtsov-PRB2019}. In the next section, we consider a quantum-mechanical model of exciton states in a QW, which is free from these approximations.    
 The model allows us to attribute the observed resonances to certain quantum-confined states of the heavy-hole excitons.  

\section{Modeling of the excitonic reflectance spectrum}

The exciton energies and wave functions for the CdTe/CdZnTe QW are calculated using numerical solution of respective Schr{\"{o}}dinger equation. The numerical procedure is described in preceding papers~\cite{Khramtsov-JAP2016, Grigoryev-SuperMicro2016, Khramtsov-PRB2019, Bataev-PRB2022}.We consider only heavy-hole exciton states and assume cylindrical symmetry of the problem, that is the anisotropy of the hole effective mass along the QW plane (the in-plane anisotropy) is ignored. This assumption reduces the problem to a three-dimensional Schr{\"{o}}dinger equation, which depends on the electron ($z_e$) and heavy-hole ($z_h$) coordinates along the growth axis and on the relative distance ($\rho$) between the electron and the hole in the QW plane. The material parameters used in the equation are listed in Tab.~\ref{Table1}. Schr{\"{o}}dinger equation is solved numerically on a $215\times 328\times 328$ regular grid (along $\rho$, $z_e$, $z_h$ coordinates) and several lowest quantum-confined exciton states are obtained.

\begin{table}[htbp!]
\caption{Material parameters for CdTe/Cd$_{0.95}$Zn$_{0.05}$Te heterostructures. $\Delta E_{g}$ is the difference of band gaps of the Cd$_{0.95}$Zn$_{0.05}$Te and CdTe; $V_e$ and $V_h$ are the band offsets for the conduction and valence bands; $m_e$ and $m_h$ are the effective masses of electron and heavy hole in CdTe; $\varepsilon$ is the dielectric constant of CdTe; $E_P$ is defined by~Eq.\ref{pcv}.
\label{Table1}}
\begin{ruledtabular}
\begin{tabular}{ccc}

& CdTe/Cd$_{0.95}$Zn$_{0.05}$Te & \\
\hline
$\Delta E_{g}$ (meV) & 27  & \cite{Kim2016,Franc2000}\\
$V_e/V_h$ & $2$\\
\hline
\hline
& CdTe & \\
\hline
$m_e$ & 0.11 $m_0$& \cite{Marple1963}\\
$m_h$ & 0.72 $m_0$& \cite{Jackson2000}\\
$\varepsilon$ & 10.2 & \cite{Perkowitz-PRB1974}\\
$E_P$ (eV)& 20.7 & \cite{Lawaetz-PRB1971,Wu-APL1985}\\
\end{tabular}
\end{ruledtabular}
\end{table}

The obtained energies of excitons states, $\hbar\omega_{0j}$, and wave functions, $\varphi_j(z_e, z_h, \rho)$, are then used to model reflectance spectra. The exciton contribution to the spectra is modeled in the framework of the nonlocal optical response theory described in a textbook by E.~L.~Ivchenko~\cite{Ivchenko-book2004}. Within this  model, the reflection from the QW is given by the sum of contributions from individual exciton resonances 
\begin{equation}
\label{Eq1}
r_{\text{QW}}(\omega)=\sum_j{\frac{i\Gamma_{0j} e^{i\phi_j}}{(\omega_{0j}+\delta\omega_{0j}-\omega)-i(\Gamma_j+\Gamma_{0j})}}.
\end{equation}
Here $\Gamma_{0j}$ describes the radiative decay rate of the $j$-th exciton state:
\begin{gather}
\label{Gamma0}
\Gamma_{0j} = \frac{2\pi q}{\hbar\varepsilon}\left(\frac{e|p_{cv}|}{m_0\omega_{0j}} \right)^2\left|\int_{-\infty}^{\infty}\Phi_j(z)\exp(i q z)\,dz \right|^2,\\
\label{pcv}
p_{cv} = \sqrt{m_0 E_p/2},
\end{gather}
where  $p_{cv}$ is the interband matrix element of the momentum operator between the electron and hole states. In~Eq.~(\ref{Gamma0})  $\Phi(z)$ is the cross-section of the exciton wave function with coinciding coordinates of the electron and the hole in the exciton, $\Phi(z) = \varphi(z, z, 0)$, $\omega_{0j}$ is the exciton resonance frequency,  $q = \sqrt{\varepsilon}\omega/c\approx \sqrt{\varepsilon} \omega_{0j}/c$ is the wave vector of light in the layer with dielectric constant $\varepsilon$,  $e$ and $m_0$ are  electron's charge and mass, respectively, $c$ is the speed of light.

The exciton-light interaction shifts the energy of bare (mechanical) exciton by some value, which is calculated as follows:
\begin{eqnarray}
&&\hbar\delta\omega_{0,j} = \frac{2\pi q}{\varepsilon}\left(\frac{e|p_{cv}|}{m_0\omega_{0j}} \right)^2 \nonumber\\
&& \hspace{3 mm}\times \iint \Phi_j(z) \Phi_j(z') \sin{\left(q \left|z-z^\prime\right|\right)} dz dz'.
\label{eqn3b}
\end{eqnarray}
The calculated shifts are given in Tab.~\ref{Table2}.

Phase $\phi_j$ entering into Eq.~(\ref{Eq1}) is calculated according to equation:
\begin{equation}
\label{phase}
\tan{\left(\frac{\phi_j}{2}\right)} = \frac{\int \Phi(z) \sin{(qz)} dz}{\int \Phi(z) \cos{(qz)} dz}\:.
\end{equation}
For the QW under consideration, $\phi_j = 0$ or $\pi$ due to the symmetry of the QW potential relative to inversion $z \rightarrow - z$.
The parameters of exciton resonances obtained in the calculations are collected in Tab.~\ref{Table2}.

\begin{table}[htbp!]
\caption{Parameters of exciton state obtained in the microscopic modeling. The band gap for CdTe, $E_g =1.606$~eV~\cite{Franc2000}. \label{Table2}}
%\begin{ruledtabular}
\begin{tabular}{l|ccccccc}
\hline
$j $ & 1 & 2 & 3 & 4 & 5 & 6 & 7 \\
\hline
$\hbar\omega_{0j}-E_g$ (meV) & -12.2 & -11.4 & -10.2 & -8.8 & -7.0 & -4.9 & -2.6\\
\hline
$\hbar\delta\omega_{0j}$ (meV) & 0.13 & -0.10 & 0.03 & 0.04 & 0.02 & -0.02 & 0.01\\

\hline
$\hbar\Gamma_{0j}$ ($\mu$eV) & 428.9 &   40.4 &  71.7 &  19.4 &  34.2 &   12.1 &   11.3\\
\hline
$\phi_{j}$ & $\pi$ & 0 & $\pi$ & 0 & $\pi$ & 0 & $\pi$ \\
\hline
\end{tabular}
%\end{ruledtabular}
\end{table}

Finally, $\Gamma_j$ is introduced into Eq.~(\ref{Eq1})  phenomenologically. It accounts for all non-radiative broadening mechanisms for the state $j$. 
In the calculations of reflectance spectra shown in Fig.~\ref{Fig1}(b), $\hbar\Gamma_j = 0.3$~meV is set for all the resonances. 
The most important mechanism of nonradiative broadening in high-quality heterostructures is the interaction of the bright excitons with other quasiparticles in the system. Note that  inhomogeneous broadening caused by imperfections of the structure as well as  nonradiative exciton recombination can, in principle,  contribute to the broadening. Optical characterization of our structure suggests, however, that the nonradiative recombination, the inhomogeneous broadening, and the structure heating (at least at the studied power densities) are small.

The total intensity of the reflected light depends also on the amplitude reflection coefficient of the sample surface, $r_s$, and can be expressed as:
\begin{equation}
   \label{Eq2}
   R(\omega) = \left|\frac{r_{s}+r_{\text{QW}}(\omega)e^{i2\theta}}{1+r_{s}r_{\text{QW}}(\omega)e^{i2\theta}}\right|^2,
\end{equation}
where $\theta$ is the phase acquired by the light wave propagating through the top layer of the structure to the middle of the QW layer. 

The reflectance spectrum calculated using Eqs.~(\ref{Eq1}, \ref{Eq2}) and data of Tab.~\ref{Table2} is shown in Fig.~\ref{Fig1}(b). To fit the experiment we have to shift  the calculated spectrum by about $2$~meV  to higher energies. The possible reason for this discrepancy between the experiment and the theory is the strain in the structure caused by the lattice constant mismatch of the QW and the barrier layers~\cite{Tuffigo-PRB1991}. The calculated spectrum is also scaled by factor $A = 0.93$. This reflects the sensitivity of the calculated values of $\hbar\Gamma_{0j}$ to model parameters, in particular, to dielectric constant $\varepsilon$, which are known with limited accuracy~\cite{Wu-APL1985, Peyla-PRB1992, Epsilon-APL1976}. Analysis shows that an increase of $\varepsilon$ in a few percents gives rise to the required  decrease of the radiative broadenings. 

With these corrections,  the calculated spectrum reproduces well the experiment. The only exception is the region between the Xhh1 and Xhh3 resonances. The observed deviation from the experimental curve is related to the light-induced mixing of exciton states. Indeed, the energy splitting between the states Xhh1, Xhh2, and Xhh3 is comparable with the light-induced shift and broadening of the states (see Tab.~\ref{Table2}). A more general (extended) model can account for this  the light-induced mixing~\cite{Voronov-FTT2007, Khramtsov-PRB2019}. The reflectance spectrum calculated using this extended model and the same set of exciton parameters (Tab.~\ref{Table2}) allowed us to obtain better agreement with the experimental results, see Fig.~\ref{Fig1}(b).

\section{Exciton dynamics}
\label{X-dynamics}

The dynamics of excitonic states was studied using  spectrally-resolved pump-probe method~\cite{Kurdyubov-PRB2021, Kurdyubov-PRB2023}. A femtosecond Ti-sapphire laser delivering $80$~fs pulses with repetition rate $\nu_L = 80$~MHz has been used in these experiments. The laser beam was divided into the pump and probe beams. The pump beam was passed through an acousto-optic tunable filter (AOTF) operating as a spectral selector. It reduced the full spectral width at the half maximum (FWHM) of the pump pulses from  $\approx 25$~meV down to $\approx 1$~meV, thus increasing the pulse duration up to $\approx 2$~ps. This spectrally narrow pump beam allowed us to selectively excite Xhh1 resonance. The spectrally broad probe beam (FWHM~$\approx 25$~meV) was used to detect reflectance spectrum at each delay between pump and probe pulses. The spectra were accumulated by the spectrometer used in the CW experiments (see Sect.~\ref{experimental}).

\begin{figure}
   \includegraphics[width=1\columnwidth]{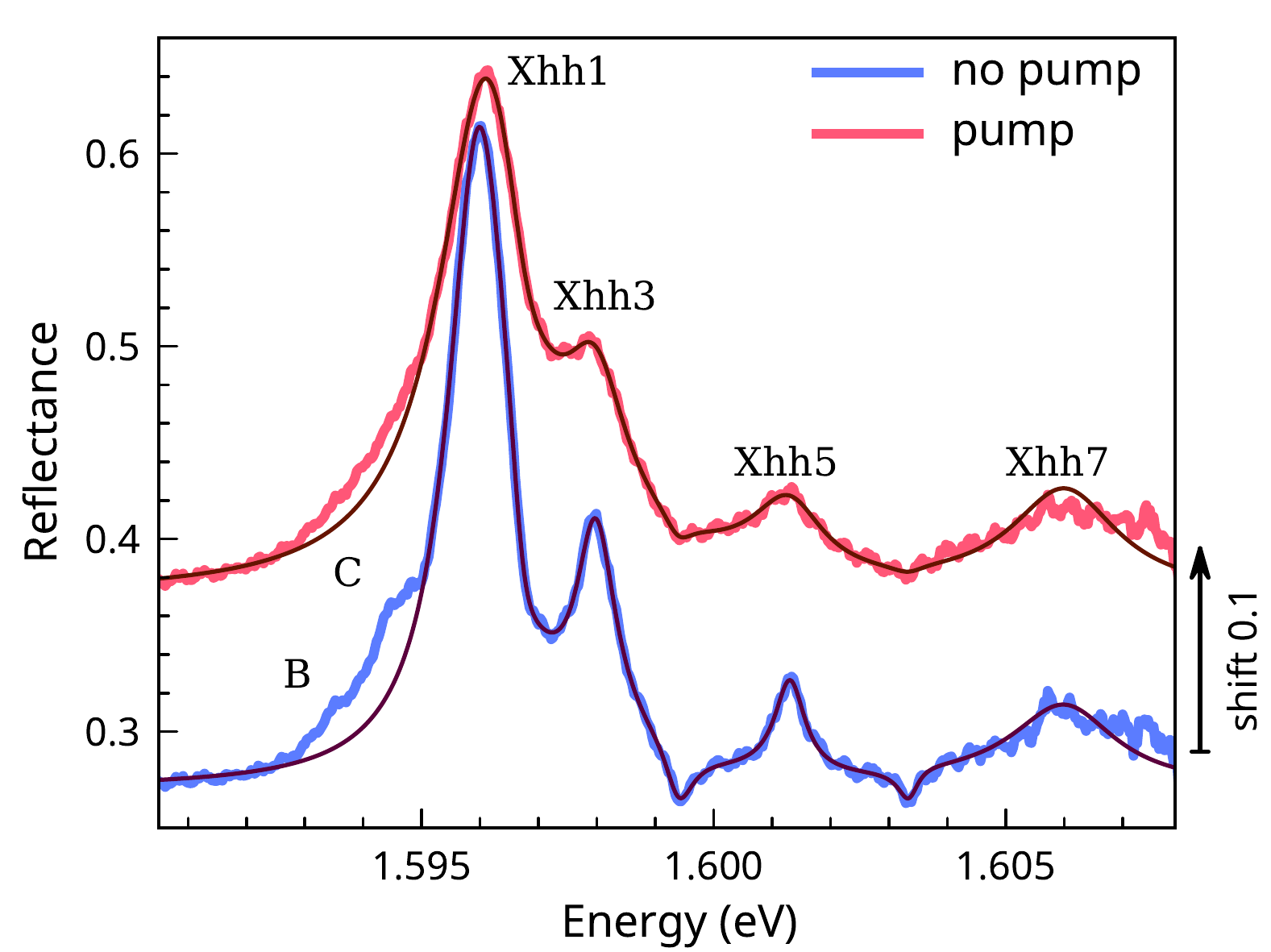}
   \caption{Exciton reflectance spectra measured with no pumping (lower thick curve) and in the presence of pumping to the Xhh1 resonance, $P_{\text{exc}} = 1.4$~mW, at the delay $\tau = 40$~ps (upper thick curve). Smooth thin curves show the fits of the spectra by Eqs.~(\ref{Eq1}, \ref{Eq2}). The upper two curves are shifted vertically by 0.1 for clarity of presentation. Sample temperature $T = 4$~K.}
   \label{Fig2}
\end{figure}

The spectra were fitted using Eqs.~(\ref{Eq1}, \ref{Eq2}). Examples of the fits are shown in Fig.~\ref{Fig2}.  The fits allow us to obtain the pump-probe delay dependence of the main excitonic parameters, namely $\hbar\omega_{0j}$, $\hbar\Gamma_{0j}$, $\hbar\Gamma_{j}$, and $\phi_j$. It appears that the exciton energies $\hbar\omega_{0j}$, the radiative broadenings $\hbar\Gamma_{0j}$, and the phases of exciton resonances $\phi_j$ are almost insensitive to the excitation of the structure with small and moderate powers used in the experiments. By contrast, the nonradiative broadenings $\hbar\Gamma_{j}$ of the exciton resonances are extremely  sensitive to the pump power and pump-probe delay. It was shown in GaAs QWs, that this photoinduced broadening contains important information about the interaction of the bright excitons with other quasiparticles in the system~\cite{Kurdyubov-PRB2021, Kurdyubov-PRB2023}. Therefore, we carefully analyze its dynamics.

\begin{figure}
   \includegraphics[width=1\columnwidth]{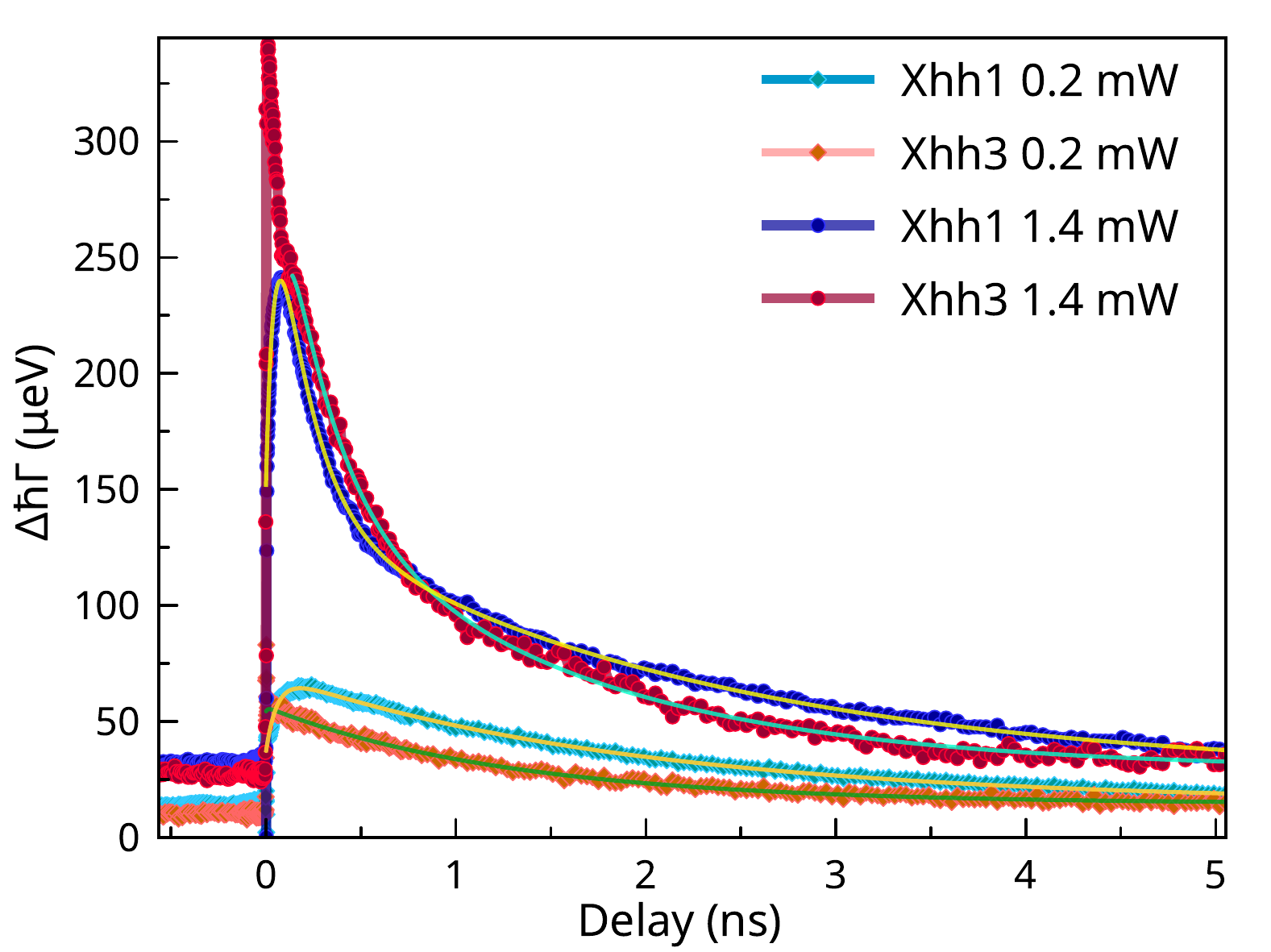}
   \caption{Dynamics of the photo-induced part of the nonradiative broadening of the resonances Xhh1 and Xhh3, powers are shown in the legend. Smooth solid curves are the fits by function~(\ref{Eq-fit}) with the time constants given in Tab.~\ref{Table3}. Sample temperature $T = 4$~K.
 }
   \label{Fig3}
\end{figure}

Figure~\ref{Fig3} shows the dependence of the pump-induced nonradiative broadening of the Xhh1 and Xhh3 resonances on the pump-probe delay for two different pump powers. The broadening not related to the pumping is subtracted from these dependencies. 
At low pumping power there is a rapid increase of the broadening, which then slowly decays with a characteristic time of about $1-2$ nanoseconds. In addition, there is a long-lived decay component that survives until the arrival of the next laser pulse. It manifests itself as a non-zero signal at negative delays (the probe pulse comes earlier than the pump pulse). This means that it  has a lifetime on the order of the pulse repetition period, $T_l = 12.5$~ns. Similar effect has been observed in  GaAs QWs~\cite{Kurdyubov-PRB2021, Kurdyubov-PRB2023}. This attests the absence of noticeable nonradiative relaxation channels, confirming the high structural quality of the sample.

At the higher pump power the photoinduced broadening increases significantly. In addition, there is another component of the broadening, which decays with a shorter characteristic time. 

The dynamics of the broadening can be approximated by a phenomenological function:
\begin{equation}
\begin{split}
	\hbar\Gamma(t) = \hbar\Gamma(-\tau_0)&\\+\Delta\hbar\Gamma_\text{max} &\cdot[c_1 e^{-t/t_1}+c_2 e^{-t/t_2} - e^{-t/t_0}],
\end{split}
\label{Eq-fit}
\end{equation}
where $-\tau_0$ is small negative delay of about -0.1~ns. The first two terms in the brackets describe the decay of the broadening with two different decay times $t_1$ and $t_2$ and the last term models the build up of the broadening just after the pump pulse with characteristic time $t_0$. The lifetime of the long-lived component of the dynamics cannot be determined with sufficient accuracy and, therefore, it is modelled as a permanent (but photo-induced) component of the broadening.

\begin{table}[htbp!]
\caption{Characteristic time constants for the fitting curves shown in Fig.~\ref{Fig3}. }
\label{Table3}
%\begin{ruledtabular}
\begin{tabular}{c|c|c|c|c}
\hline
Power~(mW) & Resonance & $t_0$~(ns) & $t_1$~(ns) & $t_2$~(ns) \\
\hline
0.2 & Xhh1 & 0.06 & -	& 1.9\\
\hline
0.2 & Xhh3 & 0.01 & -	& 1.3\\
\hline
1.4 & Xhh1 & 0.045 & 0.18 & 2.1\\
\hline
1.4 & Xhh3 & 0.045 & 0.26 & 1.4\\
\hline
\end{tabular}
%\end{ruledtabular}
\end{table}

To understand this dynamics, we start from the model developed in Refs.~\cite{Kurdyubov-PRB2021, Kurdyubov-PRB2023}. The model assumes that the long-lived broadening is caused by interaction of bright excitons with a reservoir of nonradiative excitons with large wave vectors, $K_x > K_c$. The exciton energy at the edge of the light cone is small, $E_{\text{kin}} = \hbar^2 K_c^2/(2 M_X) \approx 0.1$~meV, where $M_X = 0.9 m_0$ is the exciton translation mass along the QW layer~\cite{Dang-SSC1982}. Excitons acquire this energy at the sample temperature $T \approx 1$~K, which is much smaller than the temperature in our experiments $T=4K$. Hence, the reservoir can be easily populated by scattering of photocreated excitons with thermal phonons. In addition, the radiative broadening $\hbar\Gamma_0$ of the Xhh1 exciton state (see Tab.~\ref{Table2}) considerably exceeds $E_{\text{kin}}$ therefore the bright excitons can be scattered to nonradiative reservoir even with {\it emission} of acoustic phonons. Finally, the nonradiative excitons accumulated from previous laser pulses can also scatter the radiative excitons from the light cone. Variety of the processes resulting in the rapid population of the reservoir is observed experimentally as a fast build up of the nonradiative broadening of exciton resonaces. At the same time, this variety (also including various coherent processes) complicates the analysis of the initial stage of the dynamics. The experimentally observed  build-up of the photoinduced broadening  seems to depend on the pumping power and to be slower for Xhh1 resonance than for Xhh3. The interpretation of this result needs to be developed in the future.

The depopulation of the reservoir with characteristic times $t_1$ and $t_2$ extracted from the experiments can also occur via several processes. Here we consider possible processes only qualitatively. More rigorous analysis will be done elsewhere. 

In high-quality heterostructures at low temperatures when there is no nonradiative losses via defect centers, the only way to depopulate the reservoir is the scattering of nonradiative excitons back to the light cone where they immediately recombine. Several mechanism can be, in principle, responsible for the back scattering. These are the exciton-exciton (X-X) scattering, the exciton-electron (X-e) scattering, and the exciton-phonon (X-ph) scattering. 

The X-X scattering is, most probably, inefficient process for the reservoir depopulation, at least at moderate exciton densities realized in our experiments. Indeed, the main mechanism of the X-X interaction is the exchange interaction~\cite{Ciuti-PRB1998, Gribakin-PRB2021}. Because exciton Bohr radius in CdTe is of about $a_B\simeq5$~nm, that is significantly less than that in the GaAs ($a_B\simeq14$~nm~\cite{Khramtsov-PRB2019}), this process is expected to be even less efficient than in GaAs, where the X-e scattering dominates over X-X scattering ~\cite{Honold-PRB1989,Kurdyubov-PRB2021}.

The X-e process can be efficient when  free electrons are present in the QW and this process can explain the decay of the signal on the nanosecond scale. Some residual doping supplying the QW with electrons is probably present in the structure under study. This doping can explain the trion lines B observed in the PL spectrum, see Fig.~\ref{Fig1}(a). However, the dynamic curves measured at low excitation power do not contain any fast component, which should be present due to interaction with free electrons. We have to assume that at low power electrons are localized at some centers (donors). By contrast, stronger pumping may warm up the system, as it was verified in Ref.~\cite{Kurdyubov-PRB2023}. This results in delocalization of the electrons. Indeed, a relatively fast component of the dynamics appears  in the experiments at higher excitation power, see Fig.~\ref{Fig3}. Interaction of free electrons with the reservoir excitons results in the efficient depopulation of the reservoir. But this process does not deplete the reservoir completely, because the free electrons are cooled down and localized at the centers again. These processes control the decay of the fast component and have the characteristic time $t_1 \approx 200$~ps.

The X-phonon scattering is possibly the process controlling the slow decay components. The interaction of excitons with acoustic phonons is several times stronger in CdTe QWs~\cite{Mayer-PRB1995} than that in GaAs QWs~\cite{Rudin-PRB2002}. Our preliminary experiments show that the exciton line broadening in the PL spectra linearly increases with the sample temperature increase up to 30~K (data not shown). This increase is caused by interaction of excitons with thermal acoustic phonons and characterized by a constant $\gamma_{X-ac} \approx 7$~$\mu$eV/K describing the slope of this dependence. Similar experiments for the GaAs QWs yield $\gamma_{X-ac} = 1.5$~$\mu$eV/K~\cite{Kurdyubov-PRB2023}. The scattering of dark excitons into the light cone with emission of acoustic phonons should not be mono-exponential. As the dark reservoir is continuously cooled down, the probability of the X-photon scattering decreases due to decreasing phonon density. This could explain the presence at least two slow components in the signal, one with the characteristic time in units of nanoseconds and another one with the time exceeding the pulse repetition period. A theoretical modelling  required to accurately describe this process will be developed and presented elsewhere.

\section*{conclusion}

Our results show that multiple quantum-confined heavy-hole  exciton states contribute to the reflectivity spectrum of  wide CdTe/CdZnTe QWs.  The dynamics of the photoinduced broadening of the exciton resonances is studied under resonant pumping of the Xhh1 state. Its long-living character suggests that, as in GaAs QWs, the  photoinduced contribution to the non-radiative broadening is involves interaction with the reservoir of nonradiative (dark) excitons. At low excitation power and low sample temperature, the main processes are presumably the scattering of dark excitons to the light cone with emission of acoustic phonons followed by rapid recombination of the excitons. This is a slow process so that the reservoir is not totally depleted during the pulse repetition period of $12.5$ ns. This is evidenced by the presence of nonzero photo-induced nonradiative broadening of exciton resonances at negative time delays.

Under stronger excitation, a  faster decay component appears ($t_1\approx 200$~ps). We speculate that it  could be related to the  reservoir heating and delocalization of resident electrons, which are present in the QW under study. The X-e scattering efficiently depletes the reservoir that results in the rapid decrease of the nonradiative broadening. Simultaneously the reservoir temperature decreases, the electrons localize again, and the reservoir depletion slows down. A more rigorous discussion of the processes will be given elsewhere.

\section*{ACKNOWLEDGMENTS}
The authors acknowledge financial support from SPbU, grants No. 94030557 and No. 94271404,   and  grant ANR-21-CE30-0049 from French National Research Agency.
 I. V. I. thanks the Russian Science Foundation for the support of the theoretical part of the work (grant No. 19-72-20039). 
 B. F. G.  acknowledges support of the French Embassy in Moscow (Vernadskii fellowship for young researchers 2021). R.A. acknowledge support from the “NanoPhysics and SemiConductor” team.

 \bibliography{CdTe-v2.5-arXiv}

\end{document}